\documentclass[sn-mathphys-num]{sn-jnl}

\usepackage{graphicx}%
\usepackage{multirow}%
\usepackage{amsmath,amssymb,amsfonts}%
\usepackage{amsthm}%
\usepackage{mathrsfs}%
\usepackage[title]{appendix}%
\usepackage{xcolor}%
\usepackage{textcomp}%
\usepackage{manyfoot}%
\usepackage{booktabs}%
\usepackage{algorithm}%
\usepackage{algorithmicx}%
\usepackage{algpseudocode}%
\usepackage{listings}%
\usepackage{nicefrac}

\theoremstyle{thmstyleone}%
\theoremstyle{thmstyletwo}%

\theoremstyle{thmstylethree}%

\begin{document}

\title[Article Title]{ Annealed mean-field epidemiological model on scale-free networks with a mitigating factor }


\author*[1]{\fnm{K. M.} \sur{Kim}}

\author[1]{\fnm{M. O.} \sur{Hase}}\email{mhase@usp.br}

\affil*[1]{\orgdiv{Escola de Artes, Ci\^encias e Humanidades}, \orgname{Universidade de S\~ao Paulo}, \orgaddress{\street{Av. Arlindo B\'ettio 1000}, \city{S\~ao Paulo}, \postcode{03828-000}, \state{S\~ao Paulo}, \country{Brazil}}}


\abstract{An annealed version of the quenched mean-field model for epidemic spread is introduced and investigated analytically and assisted by numerical calculations. The interaction between individuals follows a prescription that is used to generate a scale-free network, and we have adjusted the number of connections to produce a sparse network. Specifically, the model's behavior near the infection threshold is examined, as well as the behavior of the stationary prevalence and the probability that a connection between individuals encounters an infected one. We found that these functions display a monotonically increasing dependence on the infection rate. Subsequently, a modification that mimics the mitigation in the probability of encountering an infected individual is introduced, following an old idea rooted in the Malthus-Verhulst model. We found that this modification drastically changes the probability that a connection meets an infected individual. However, despite this change, it does not alter the monotonically increasing behavior of the stationary prevalence.
}




\maketitle


\section{Introduction}
\label{sec:introduction}

Infectious diseases have historically been a major cause of mortality, especially in low-income countries \cite{WHO}. The recent COVID-19 pandemic has underscored the potential for infectious diseases to emerge and spread on a global scale. Forecasting the spread of such diseases presents significant challenges, exacerbated by incomplete data due to underreporting and the ethical constraints of conducting experiments on disease spread in human populations. Consequently, mathematical modeling has emerged as a vital tool for addressing these challenges \cite{H89}.

In epidemiology, mathematical modeling often involves compartmentalizing the population into distinct states regarding the disease, such as susceptible, infected, recovered/removed, and others. The fundamental approach is to develop a set of differential equations that describe the transitions of individuals between these compartments \cite{M02, KR08, TdO20}. A crucial aspect of these models is comprehending the interaction network among individuals, particularly when the disease is transmitted through direct contact. Traditional models frequently assume \textquotedblleft homogeneous mixing\textquotedblright, where each individual has an equal probability of interacting with any other individual \cite{H06}. Although this approach has been widely adopted for decades \cite{B75}, it fails to accurately represent the heterogeneous nature of human contact networks \cite{DM03}. Therefore, there is a pressing need for a theory that combines classical mathematical epidemiology \cite{AM92, M02} with network theory \cite{AB02, N10, dM20, DM22} to enhance our understanding of disease spread in a more realistic context \cite{FCPS12, PSCvMV15}.

There are two important mean-field models for incorporating the contact network between individuals in the context of epidemiology. In the heterogeneous mean-field model (HMF) \cite{PSV01a, PSV01b}, vertices are partitioned based on their degrees: nodes with the same number of connections are considered statistically equivalent, and the contact network information is incorporated through the degree distribution. This scenario typically assumes that the network's evolution is much faster than any dynamical process occurring on it; in other words, these processes effectively take place on an average network resulting from the rewiring, under the constraint of maintaining the degree distribution \cite{PSCvMV15}. In the context of disordered systems \cite{MPV87}, this is known as the annealed limit. Conversely, in the quenched limit, the network structure is fixed -- at least on the timescale of the dynamical processes occurring within the system \cite{MPV87}. In this situation, since the contact network configuration is fixed, the network information is introduced through its adjacency matrix \cite{HY84, WCWF03}, constituting the quenched mean-field approach. It is well known that these two different timescales generally lead to different scenarios \cite{HM08, CH16}.

The main goal of this work is twofold. First, we aim to investigate the effect of introducing the annealed approximation to an epidemic model originally in its quenched version. The details of this model are provided below, but it can be seen as a step toward the quenched timescale, as the adjacency matrix is annealized without initially referencing the degree distribution like in HMF (although it can be determined subsequently). Secondly, we aim to examine an epidemic model where the state of the vertices influences individual behavior. Specifically, we are interested in a situation where infected nodes may effectively lose some contact with other individuals due to voluntary social isolation and/or hospitalization. For instance, one can weaken the ties to an infected individual by tuning a parameter that controls the probability of interaction between individuals \cite{DH21}. Another possibility, explored in \cite{KDH23}, is to invoke an old idea that goes back to the classical work of Verhulst \cite{V38}. In population dynamics, the Malthusian model \cite{M98} posits that the growth rate of a population is proportional to its size, leading to exponential growth in the absence of constraints and assuming an infinite supply of resources. To prevent unchecked growth, Verhulst introduced the concept of carrying capacity, which limits the maximum population size. He replaced the linear growth term in the Malthusian model with a parabolic term with negative convexity, leading to a logistic growth function. This simple modification has profound implications, as population growth follows a logistic curve. This concept has been previously applied in the context of networks \cite{HCG16}, and we are now extending it to describe the dynamics of individual interactions during an epidemic.

In \cite{KDH23}, this effect was introduced into the heterogeneous mean-field model \cite{PSV01a, PSV01b} defined on a Barab\'asi-Albert network \cite{BA99}. It was shown that the prevalence was a monotonically increasing function of the infection rate, although the probability of a connection encountering an infected individual did not behave accordingly, displaying a maximum. This difference between these two quantities is not observed in the original HMF model, where both functions increase with the infection rate. We aim to explore this in the annealed quenched model to examine how the network connection is accessed.

Naturally, these are all mean-field models, and the aim is not to focus on universal quantities like critical exponents, but rather on non-universal ones like the critical point (infection threshold), amplitudes of the order parameters, and other qualitative changes of some key quantities involved in the analysis of the system.

The layout of this work is as follows. In Section 2, we review the two main mean-field approaches to epidemic models that incorporate the network structure of contacts between individuals: the heterogeneous and quenched mean-field models. In Section 3, we introduce the annealed version of the quenched mean-field model, followed by its modified version in the subsequent section. The conclusions and discussion of the results are summarized in the final section.


\section{Annealed and quenched mean-field models}
\label{sec:hmfqmf}

We will begin by briefly revisiting two mean-field models applied to SIS dynamics. In the HMF \cite{PSV01a, PSV01b} approach, the individuals in the population are divided according to their degrees, with those sharing the same degree considered statistically similar. This contrasts with the traditional strategy based on the homogeneous mixing hypothesis \cite{H06, KR08}, where all individuals are treated equally, leading to a description of a homogeneous network. In such a network, all vertices have similar degrees, and those with significantly more (or fewer) connections than the mean value are rare. The HMF formulation, on the other hand, allows for the investigation of heterogeneous networks, as shown below. The master equation for this model is given by:
\begin{align}
\frac{\textup{d}}{\textup{d}t}\rho_{k}(t) = -\rho_{k}(t) + \lambda\big[1-\rho_{k}(t)\big]k\Theta_{k}(t),
\label{hmf}
\end{align}
where $\rho_{k}(t)$ is the density of infected individuals with degree $k$ at time $t$, and $\Theta_{k}(t)$ represents the probability that a link emanating from a node with degree $k$ connects to an infected node at time $t$. The time scale has been adjusted so that the recovery rate is fixed at $1$, and the infection rate is $\lambda$ in this setup. The probability $\Theta_{k}(t)$ can be cast as:
\begin{align}
\Theta_{k}(t) = \sum_{k^{\prime}}P(k^{\prime}|k)\rho_{k^{\prime}}(t),
\label{Thetak}
\end{align}
where $P(k_{1}|k_{2})$ denotes the conditional probability that a link from a node with degree $k_{2}$ connects to a vertex with degree $k_{1}$. The main characterization of the underlying network is assigned to this conditional probability. In the case of an uncorrelated network, where the degrees of the two vertices at either end of an edge are independent, it is known that the infection threshold is given by \cite{BPSV03}
\begin{align}
\lambda_{c}^{hmf} = \frac{\langle k\rangle}{\langle k^{2}\rangle},
\label{lambdac_nc}
\end{align}
where
\begin{align}
\langle k^{n}\rangle=\sum_{k}k^{n}P(k)
\label{nmoment}
\end{align}
is the $n$-th moment. This indicates that, for SIS dynamics on an uncorrelated network, the stationary regime is disease-free if $\lambda<\lambda_{c}$, while the infection persists if $\lambda>\lambda_{c}$. In a homogeneous network where all nodes are considered to have essentially the same degree $\langle k\rangle$, the critical point is reduced to $\lambda_{c}=1/\langle k\rangle$. On the other hand, for an infinite scale-free network following a power-law degree distribution $P(k)\sim k^{-\gamma}$, the infection threshold is
\begin{align}
\lambda_{c}^{hmf} = \left\{
\begin{array}{ccl}
  0 &,& \gamma\leq 3 \\
\displaystyle\left(\frac{\gamma-3}{\gamma-2}\right)\frac{1}{m} &,& \gamma>3
\end{array}
\right.,
\label{lambdac_hmf_sf}
\end{align}
where $m$ is the minimum degree of the network. It is noteworthy that, when a power-law function is used to describe the degree distribution, the corresponding exponents for many systems are typically smaller than $3$ \cite{N03, CSN09}. Moreover, the prevalence (the stationary density of the infected population) $\rho_{\infty}$ for the HMF can also be characterized on an uncorrelated scale-free network \cite{PSV01b, DH21} as
\begin{align}
\rho_{\infty}^{hmf} \sim \left\{
\begin{array}{ccl}
\displaystyle\lambda^{\frac{1}{3-\gamma}} &,& 2<\gamma< 3 \\
\displaystyle\left(\lambda-\lambda_{c}\right)^{\frac{1}{\gamma-3}} &,& 3<\gamma<4 \\
\displaystyle\left(\lambda-\lambda_{c}\right)^{1} &,& \gamma>4
\end{array}
\right..
\label{prevalence_hmf_sf}
\end{align}

Another important approach to the SIS dynamics on mean-field level is the quenched mean-field model \cite{HY84, WCWF03}, whose master equation, 
\begin{align}
\frac{\textup{d}}{\textup{d}t}\rho_{i}(t) = -\rho_{i}(t) + \lambda\big[1-\rho_{i}(t)\big]\sum_{j}A_{ij}\rho_{j}(t),
\label{qmf}
\end{align}
describes the time evolution of $\rho_{i}(t)$, which is the probability that node $i$ is infected at time $t$. Once again, the time scale has been set so that the recovering rate is $1$. Information about the underlying contact network between individuals is encoded in the adjacency matrix $(A_{ij})$, whose elements are defined by
\begin{align}
A_{ij}=\left\{
\begin{array}{ccl}
1 &,& \text{there is a connection between vertices $i$ and $j$} \\
0 &,& \text{there is no connection between vertices $i$ and $j$}
\end{array}
\right..
\label{Aij}
\end{align}
The linear stability analysis on this model indicates that the infection threshold is given by \cite{CPS10}
\begin{align}
\lambda_{c}^{qmf} = \frac{1}{\Lambda_{\max}},
\label{lambdac_qmf_sf}
\end{align}
where $\Lambda_{\max}$ is the largest eigenvalue of the adjacency matrix. For a scale-free network, it is known that \cite{CLV03} $\Lambda_{\max}\sim\max\{\sqrt{k_{\max}},\nicefrac{\langle k^{2}\rangle}{\langle k\rangle}\}$, where $k_{\max}$ is the largest degree of the network. Therefore, one can then see that in the thermodynamic limit \cite{CD09},
\begin{align}
\lambda_{c}^{qmf} = 0.
\label{lambdac=0_qmf_sf}
\end{align}
Nevertheless, a deeper analysis of the SIS dynamics reveals certain inconsistencies. These include the absence of a critical threshold when the principal eigenvector of the adjacency matrix is localized, as observed for $\gamma > 5/2$ in uncorrelated scale-free networks \cite{GDOM12}, and the presence of a nonzero infection threshold when the dynamics consider the interplay between two large hubs in contact with each other \cite{LSN13}. Furthermore, extending beyond mean-field analysis highlights the critical role of interactions between hubs -- whether or not they are in close proximity -- in driving the mechanism of epidemic spread \cite{BCPS13, CPS20}. This approach ultimately leads to the expression \eqref{lambdac_qmf_sf} in the thermodynamic limit.




\section{Annealed mean-field model}
\label{sec:aqmf}

The differences between the annealed and quenched approach to SIS dynamics on mean-field level were briefly discussed in the previous section. As it is known, in the network literature, these concepts are related to different time scales of the degree random variable. The annealed case can deal with a situation where the links are reconfigured several times during the observation period, and the information about the structure of the graph of contacts between individuals is introduced by its mean behavior. This can be accomplished by using the degree distribution of the system. On the other hand, when the configuration of the network is fixed during the observation time, the adjacency matrix is the natural quantity to incorporate in the model \cite{PSCvMV15}.

In this work, we propose an alternative approach where annealing is introduced on the model from the quenched mean-field model \eqref{qmf} by relaxing the adjacency matrix to its annealed version. In other words, we replace the element $A_{ij}$, which indicates whether nodes $i$ and $j$ are connected, by $p_{ij}$, which stands for the probability that these vertices are linked.

We will always consider a scale-free network whose degree distribution is given by
\begin{align}
P(k) = \left(\gamma-1\right)m^{\gamma-1}k^{-\gamma}\qquad (k\geq m),
\label{Psf}
\end{align}
where $m$ is the minimum degree of the network. Here, we adopt the continuous approximation, where degrees are treated as continuous variables. One can easily verify the normalization condition for \eqref{Psf} and the mean degree
\begin{align}
\langle k\rangle_{SF} = \left(\frac{\gamma-1}{\gamma-2}\right)m.
\label{<k>Psf}
\end{align}
In principle, the condition $\gamma>1$ is sufficient to ensure a \textit{bona fide} degree distribution for an infinite network, but if we also demand a finite mean degree, we have to admit that $\gamma>2$.


\subsection{Disconnected network}
\label{subsec:disconnected}

The results presented in this subsection are straighforward and not particularly noteworthy \textit{per se}, but they serve a twofold purpose: (i) to introduce some notations and (ii) to establish a base case that helps us in the construction of the sparse network case discussed in the next subsection.

For a network with $N$ vertices, our starting point is going to be the master equation

\begin{align}
\frac{\textup{d}}{\textup{d}t}\rho_{i}(t) = -\rho_{i}(t) + \lambda\big[1-\rho_{i}(t)\big]\sum_{j}\tilde p_{ij}\rho_{j}(t),
\label{aqmf}
\end{align}
where the adjacency matrix $(A_{ij})$ in \eqref{qmf} was replaced by $(\tilde p_{ij})$, which is the probability that the nodes $i$ and $j$ are linked. Although at this point the choice of $\tilde p_{ij}$ is arbitrary, we first assume that
\begin{align}
\tilde p_{ij} = \frac{i^{-\alpha}j^{-\alpha}}{Z_{\alpha}},
\label{pij}
\end{align}
where
\begin{align}
Z_{\alpha} := \sum_{i=1}^{N}i^{-\alpha},
\label{Zalpha}
\end{align}
since this choice leads to a scale-free network, consistent with known algorithms for constructing networks that obey a power-law degree distribution \cite{GKK01, CLV03}. One can assign weights to vertices such that the linkinig probability is given by \eqref{pij}, and the relation between the exponent $\alpha$, associated with these weights (see \eqref{pij}), and the exponent $\gamma$, from the degree distribution \eqref{Psf}, is given by \cite{GKK01}
\begin{align}
\alpha = \frac{1}{\gamma-1}.
\label{alphagamma}
\end{align}

The mean degree of the vertex $i$ in this model is
\begin{align}
\langle\tilde k_{i}\rangle := \sum_{j=1}^{N}\tilde p_{ij} = i^{-\alpha},
\label{ki}
\end{align}
and recovers a uniform network when $\alpha=0$. It is clear from \eqref{ki} that the mean degree is (equal to or) smaller than $1$, indicating that the graph is disconnected. Nevertheless, we will explore this scenario before fixing this undesirable property (from the standpoint of not being comparable with real network of contact between individuals) in the next section.

As expected, the trivial stationary solution $\vec\rho(\infty)=(0,\ldots,0)$ is a solution of \eqref{aqmf}, and the stability analysis follows the prescription shown in section \ref{sec:hmfqmf}. We have to analyze the largest eigenvalue of the matrix $(\tilde p_{ij})$, which is $Z_{2\alpha}/Z_{\alpha}$, which leads to the infection threshold
\begin{align}
\lambda_{c}^{dis} = \frac{Z_{\alpha}}{Z_{2\alpha}} \simeq\left\{
\begin{array}{lcll}
\displaystyle\frac{1-2\alpha}{1-\alpha}N^{\alpha} &,& 0\leq\alpha<\nicefrac{1}{2} & (\gamma>3) \\
 & & & \\
\displaystyle 2\frac{\sqrt{N}}{\ln N} &,& \alpha=\nicefrac{1}{2} & (\gamma=3) \\ 
 & & & \\
\displaystyle \frac{1}{\left(1-\alpha\right)\zeta(2\alpha)}N^{1-\alpha} &,& \nicefrac{1}{2}<\alpha<1 & (2<\gamma<3) \\ 
 & & & \\
\displaystyle \frac{1}{\zeta(2)}\ln N &,& \alpha=1 & (\gamma=2) \\ 
 & & & \\
\displaystyle \frac{\zeta(\alpha)}{\zeta(2\alpha)} &,& \alpha>1 & (\gamma<2)
\end{array}
\right.,
\label{lambdac_aqmf}
\end{align}
where
\begin{align}
\zeta(\nu) := \sum_{i=1}^{\infty}i^{-\nu} \quad (\nu>1)
\label{zeta}
\end{align}
is the usual zeta function. The asymptotic behavior with respect to the size of the system in \eqref{lambdac_aqmf} shows that the critical point is finite for $\gamma<2$ only. In the other cases, the infection threshold diverges in the thermodynamic limit, meaning the system is confined to the disease-free absorbing state. This scenario is expected since the network is disconnected and we do not anticipate disease prevalence. On the other hand, the case $\gamma<2$ is special. As discussed earlier in this section, the minimum requirement for the exponent $\gamma$ of the degree distribution is $\gamma>1$. When $1<\gamma<2$, the degree distribution is so heavily-tailed that a tiny fraction of the nodes have extremely high degrees, playing a central role in preserving the infection even in a graph below the percolation threshold. From \eqref{zeta}, it is evident that the extremal case $\alpha\rightarrow\infty$ (or $\gamma\rightarrow 1^{+}$) leads to $\lambda_{c}^{dis}\rightarrow 1$.

The divergence of the critical point \eqref{lambdac_aqmf} implies that the system is confined to an absorbing state in the thermodynamic limit (the only exception is the case $\alpha>1$, which will soon be excluded from analysis). In other words, the prevalence (the stationary fraction of infected population) is alwawys zero irrespective of the infection rate. This anomalous situation is a consequence of the low connectivity of the graph, which is disconnected, as expected from the small mean degree \eqref{ki} and pointed out before. We will address this issue in the next section.


\subsection{Sparse network}
\label{subsec:sparse}

The previously introduced model suffered from the issue of generating a disconnected graph, which is undesirable for investigating disease propagation. We will now address this issue through a simple procedure. Initially, the (disconnected) network was generated by assigning the probability \eqref{pij} to link two nodes. Since this approach resulted in a graph with a low number of edges, we can mitigate this by repeating the linking process multiple times. Specifically, we attempt to link each pair of vertices $\Omega$ times to achieve a connected network. While the choice of $\Omega$ is arbitrary (as long as it is sufficiently large), it can be selected so that the mean degree of the now connected network matches \eqref{<k>Psf}. Although it may not be possible to ensure that the minimum degree of the resulting graph with $N$ vertices is always $m$, the total number of links will be $\langle k\rangle_{SF} N$, corresponding to a sparse network. From \eqref{ki}, the total degree of the disconnected network is $Z_{\alpha}$, given by \eqref{Zalpha}, which requires calibrating $\Omega$ such that
\begin{align}
Z_{\alpha}\Omega = \langle k\rangle_{SF} N = \left(\frac{\gamma-1}{\gamma-2}\right)mN \quad\textnormal{ or }\quad \Omega = \frac{mN}{\left(1-\alpha\right)Z_{\alpha}},
\label{Omega}
\end{align}
where \eqref{alphagamma} was also invoked. We are now requiring $\alpha<1$ ($\gamma>2$) to have a well-behaved mean degree.

The master equation for this sparse network is then given by
\begin{align}
\frac{\textup{d}}{\textup{d}t}\rho_{i}(t) = -\rho_{i}(t) + \lambda\big[1-\rho_{i}(t)\big]\sum_{j}p_{ij}\rho_{j}(t),
\label{aqmf_ext}
\end{align}
where
\begin{align}
p_{ij} := \Omega\tilde p_{ij}.
\label{ptilde}
\end{align}
A related work, where the annealed adjacency matrix was based on the degrees (instead of the node labels) was examined in \cite{KJI06}.

The linear stability analysis of \eqref{aqmf_ext} leads to an infection threshold $\lambda_{c}=\Omega^{-1}\lambda_{c}^{dis}$, which is
\begin{align}
\lambda_{c} = \left(\frac{1-\alpha}{mN}\right)\frac{Z_{\alpha}^{2}}{Z_{2\alpha}}\simeq \left\{
\begin{array}{lcll}
\displaystyle\left(\frac{1-2\alpha}{1-\alpha}\right)\frac{1}{m} &,& 0\leq\alpha<\nicefrac{1}{2} & (\gamma>3) \\
 & & & \\
\displaystyle\frac{2}{m}\frac{1}{\ln N} &,& \alpha = \nicefrac{1}{2} & (\gamma=3) \\
 & & & \\
\displaystyle\frac{1}{\left(1-\alpha\right)\zeta(2\alpha)}\frac{1}{m}N^{-\left(2\alpha-1\right)} &,& \nicefrac{1}{2}<\alpha<1 & (2<\gamma<3)
\end{array}
\right..
\label{lambdac_aqmf_sparse}
\end{align}
From \eqref{lambdac_aqmf_sparse}, we now have a new scenario, where the system exhibits a finite critical point $\lambda_{c}$ that separates the active state from the absorbent phase if $0\leq\alpha<\nicefrac{1}{2}$ ($\gamma>3$), while this infection threshold vanishes in the thermodynamic limit if $\nicefrac{1}{2}\leq\alpha<1$ ($2<\gamma\leq 3$).

Next, we investigate the stationary regime. In this case, the infection probability of site $i$ no longer varies with time, i.e., $\frac{\textup{d}\rho_{i}}{\textup{d}t}=0$, and consequently, $\rho_{i}(t)=\rho_{i}$ for $1\leq i\leq N$ becomes independent of time. The main quantity describing the behavior of the system is the prevalence $\rho$, which is the stationary fraction of the infected population, given by
\begin{align}
\rho := \frac{1}{N}\sum_{i=1}^{N}\rho_{i}.
\label{prevalence}
\end{align}

The analysis below begins with the stationary equation
\begin{align}
\rho_{i} = \lambda\left(1-\rho_{i}\right)\sum_{j}p_{ij}\rho_{j},
\label{pre_basic_aqmfext}
\end{align}
which follows directly from \eqref{aqmf_ext}. From \eqref{Omega}, \eqref{ptilde} and \eqref{lambdac_aqmf_sparse}, we have
\begin{align}
\rho_{i} = \frac{\displaystyle\frac{\lambda}{\lambda_{c}}\frac{i^{-\alpha}}{Z_{2\alpha}}\psi_{\alpha}}{1 + \displaystyle\frac{\lambda}{\lambda_{c}}\frac{i^{-\alpha}}{Z_{2\alpha}}\psi_{\alpha}},
\label{basic_aqmfext}
\end{align}
where
\begin{align}
\psi_{\alpha} := \sum_{i=1}^{N}i^{-\alpha}\rho_{i}
\label{psi}
\end{align}
depends on the infection rate $\lambda$. Multiplying both sides of \eqref{basic_aqmfext} by $i^{-\alpha}$ and summing over $i$ gives
\begin{align}
\psi_{\alpha}(\lambda) = \frac{\lambda}{\lambda_{c}}\frac{\psi_{\alpha}}{Z_{2\alpha}}I_{2,1}(\lambda), \quad \text{ where } \quad I_{m,n}(\lambda) := \sum_{i}\frac{i^{-m\alpha}}{\left[1 + \frac{\lambda}{\lambda_{c}}\frac{i^{-\alpha}}{Z_{2\alpha}}\psi_{\alpha}(\lambda)\right]^{n}}.
\label{I}
\end{align}
For $m,n>0$, the following inequalities hold:
\begin{align}
I_{a,n}(\lambda) < I_{b,n}(\lambda) \quad\text{ for any $\lambda$ and $a>b$}, \quad\text{ and }\quad I_{m,c}(\lambda) < I_{m,d}(\lambda) \quad\text{ if $\lambda>\lambda_{c}$ and $c>d$}.
\label{I<I}
\end{align}

Let us now investigate the supercritical region close to the transition point $\lambda_{c}$. In this regime, \eqref{basic_aqmfext} leads to
\begin{align}
\left\{
\begin{array}{lcl}
N\rho &=& \displaystyle\frac{\lambda}{\lambda_{c}}\frac{Z_{\alpha}}{Z_{2\alpha}}\psi_{\alpha} - \left(\frac{\lambda}{\lambda_{c}}\right)^{2}\frac{1}{Z_{2\alpha}}\psi_{\alpha}^{2} + \cdots \\
\psi_{\alpha} &=& \displaystyle\frac{\lambda}{\lambda_{c}}\psi_{\alpha} - \left(\frac{\lambda}{\lambda_{c}}\right)^{2}\frac{Z_{3\alpha}}{Z_{2\alpha}^{2}}\psi_{\alpha}^{2} + \cdots
\end{array}
\right.
\label{sp_exp}
\end{align}
by keeping terms up to the second order. From \eqref{sp_exp}, one sees that
\begin{align}
\psi_{\alpha} \simeq \frac{Z_{2\alpha}^{2}}{Z_{3\alpha}\lambda_{c}}\left(\lambda - \lambda_{c}\right) \quad\text{ and }\quad N\rho \simeq \frac{Z_{\alpha}Z_{2\alpha}}{Z_{3\alpha}\lambda_{c}}\left(\lambda - \lambda_{c}\right),
\label{sp_rhopsi}
\end{align}
which is consistent with our analysis at the mean-field level.

Finally, we can estimate the probability $\Theta$ of a link meeting an infected node, \eqref{Thetak}, within the context of this annealed mean-field model. The natural way to define this quantity is through
\begin{align}
\Theta = \frac{1}{N}\sum_{i}\frac{1}{\langle k_{i}\rangle}\sum_{j}p_{ij}\rho_{j} = \frac{\psi_{\alpha}}{Z_{\alpha}},
\label{Theta_annq}
\end{align}
where $\langle k_{i}\rangle=\sum_{j}p_{ij}=\Omega i^{-\alpha}$ is the mean degree of vertex $i$.

\begin{figure}
\centering
\includegraphics[width=368pt]{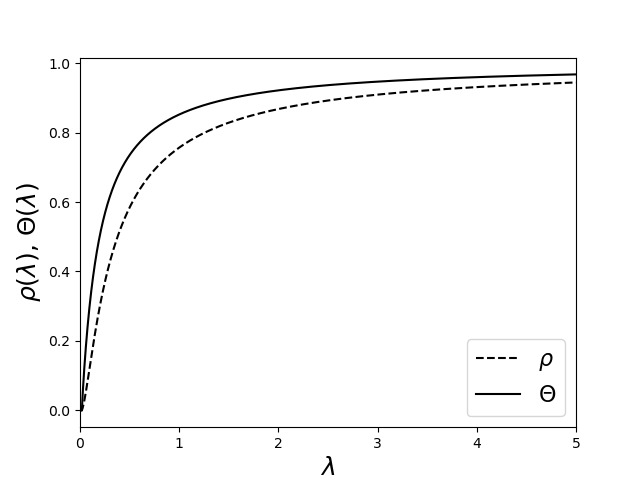}
\caption{Graph $\rho\times\lambda$ and $\Theta\times\lambda$ in the annealed mean-field model; boh functions are increasing in the supercritical regime. The parameters are $m=2$, $\gamma=\nicefrac{5}{2}$ ($\alpha=\nicefrac{2}{3}$) and $N=10^{4}$. The critical point is $\lambda_{c}=0.01933\ldots$, and the subcritical absorbent phase is not visible in the graph.}
\label{sparsegraph}
\end{figure}

The probability \eqref{Theta_annq} is a non-decreasing function with respect to the infection rate. Assuming that $\lambda>\lambda_{c}$ (which implies $\psi_{\alpha}\neq 0$), differentiating both sides of \eqref{I} with respect to $\lambda$ yields
\begin{align}
\frac{\textup{d}}{\textup{d}\lambda}\psi_{\alpha} = \frac{\psi_{\alpha}I_{2,2}(\lambda)}{\lambda\Big[ I_{2,1}(\lambda) - I_{2,2}(\lambda) \Big]}
\label{dpsidlambda}
\end{align}
after some tedious manipulations. Since $I_{2,1}(\lambda) > I_{2,2}(\lambda)$ from \eqref{I<I} in the supercritical regime, one sees that the derivative of the probability \eqref{Theta_annq} is positive in the supercritical regime. Furthermore, from \eqref{basic_aqmfext}, one sees that
\begin{align}
\frac{\textup{d}}{\textup{d}\lambda}\left(N\rho\right) = \frac{\psi_{\alpha} + \lambda\psi_{\alpha}^{\prime}}{\lambda_{c}Z_{2\alpha}}I_{1,2}(\lambda) > 0.
\label{drhodlambda}
\end{align}
Therefore, both the prevalence and the probability \eqref{Theta_annq} are monotonically increasing functions in the supercritical regime. This behavior is illustrated in the graph \ref{sparsegraph}.

\section{Modified annealed mean-field model}
\label{sec:maqmf}

We now introduce a modification to the infection process by incorporating a mitigating factor. This factor simulates the tendency of infected individuals to reduce their participation in the contact network. For this analysis, we will focus solely on the model defined on a sparse network. The master equation for this modified model is given by
\begin{align}
\frac{\textup{d}}{\textup{d}t}\rho_{i}(t) = -\rho_{i}(t) + \lambda\big[1-\rho_{i}(t)\big]\sum_{j}p_{ij}\rho_{j}(t)\big[1-\rho_{j}(t)\big],
\label{mme}
\end{align}
and the relation
\begin{align}
\rho_{i} = \lambda\big(1-\rho_{i}\big)\sum_{j}p_{ij}\rho_{j}\big(1-\rho_{j}\big),
\label{mme_stationary}
\end{align}
is valid in the stationary regime.

Note that by this modification, the probability $\Theta_{m}$ (here, the subscript $m$ stands for the modified model, and there should be no confusion with the degree variable, as in \eqref{Thetak}, since under the uncorrelated network approximation, such degree dependence was no longer taken into account) of a link meeting an infected node is cast as
\begin{align}
\Theta_{m} = \frac{1}{N}\sum_{i}\frac{1}{\langle k_{i}\rangle}\sum_{j}p_{ij}\rho_{j}\left(1-\rho_{j}\right) = \frac{\psi_{\alpha}-\phi_{\alpha}}{Z_{\alpha}},
\label{mTheta}
\end{align}
where
\begin{align}
\phi_{a} := \sum_{i=1}^{N}i^{-\alpha}\rho_{i}^{2}.
\label{phia}
\end{align}
The probability \eqref{mTheta} is clearly lower than the corresponding value in the original model, as given by \eqref{Theta_annq}. Furthermore, since $\rho_{j}\left(1-\rho_{j}\right)\leq\frac{1}{4}$ for $0\leq\rho_{j}\leq 1$, there is a natural bound
\begin{align}
\Theta_{m}\leq \frac{1}{4}.
\label{Thetam<1/4}
\end{align}

Linear stability analysis reveals that the critical point remains unchanged from the original model, so we will continue to use the symbol $\lambda_{c}$ to represent the infection threshold in this modified model. In the stationary regime, equation \eqref{mme} leads to
\begin{align}
\rho_{i} = \frac{\lambda}{\lambda_{c}}\frac{i^{-\alpha}}{Z_{2\alpha}}\left(1-\rho_{i}\right)\left(\psi_{\alpha}-\phi_{\alpha}\right),
\label{mhmf_st}
\end{align}
and from \eqref{mhmf_st}, one has
\begin{align}
\psi_{\alpha} = \frac{\lambda}{\lambda_{c}}\frac{\left(Z_{2\alpha}-\psi_{2\alpha}\right)}{Z_{2\alpha}}\left(\psi_{\alpha}-\phi_{\alpha}\right).
\label{mhmfpsi}
\end{align}

Next, we analyze the behavior near the critical point $\lambda\sim\lambda_{c}^{+}$. Following a similar procedure as in the previous section, we obtain the relations
\begin{align}
\left\{
\begin{array}{lcl}
N\rho &=& \displaystyle\frac{\lambda}{\lambda_{c}}\frac{Z_{\alpha}}{Z_{2\alpha}}\left(\psi_{\alpha} - \phi_{\alpha}\right) - \left(\frac{\lambda}{\lambda_{c}}\right)^{2}\frac{1}{Z_{2\alpha}}\psi_{\alpha}^{2} + \cdots \\
\psi_{\alpha} &=& \displaystyle\frac{\lambda}{\lambda_{c}}\left(\psi_{\alpha} - \phi_{\alpha}\right) - \left(\frac{\lambda}{\lambda_{c} Z_{2\alpha}}\right)^{2}\psi_{\alpha}^{2} + \cdots \\
\phi_{\alpha} &=& \displaystyle\frac{\lambda}{\lambda_{c}}\frac{\psi_{2\alpha}}{Z_{2\alpha}}\psi_{\alpha} + \cdots
\end{array}
\right.
\label{mhmf_exp}
\end{align}
from \eqref{mhmf_st}. Then, from \eqref{mhmfpsi} and \eqref{mhmf_exp}, one has
\begin{align}
\psi_{\alpha} \simeq \frac{Z_{\alpha}^{2}}{2\lambda_{c}}\left(\lambda-\lambda_{c}\right) \quad\text{ and }\quad N\rho \simeq \frac{Z_{\alpha}Z_{2\alpha}}{\lambda_{c}}\left(\lambda-\lambda_{c}\right) \quad\text{ for }\quad \lambda\sim\lambda_{c}^{+}.
\label{mhmf_criticality}
\end{align}
Although the coefficients (amplitudes) in \eqref{mhmf_criticality} differ from those in \eqref{sp_rhopsi}, both exhibit the same mean-field exponent.

\begin{figure}
\centering
\includegraphics[width=368pt]{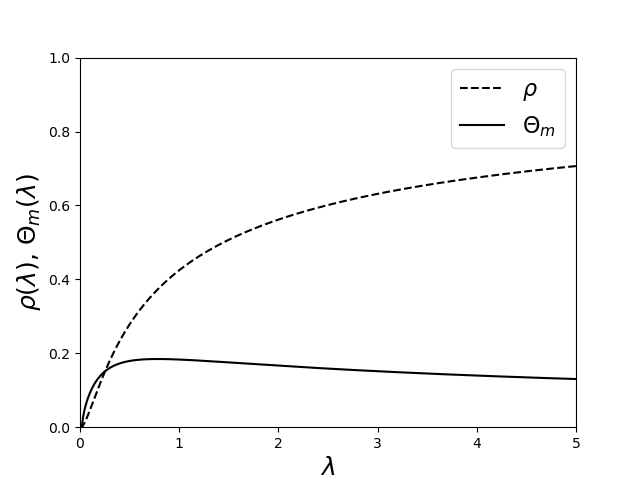}
\caption{Graph $\rho\times\lambda$ and $\Theta_{m}\times\lambda$ for $m=2$, $\gamma=\nicefrac{5}{2}$ ($\alpha=\nicefrac{2}{3}$) and $N=10^{4}$.}
\label{mhmffigure}
\end{figure}

As one can see from figure \ref{mhmffigure}, although the prevalence increases with the infection rate $\lambda$ (like in the annealed model), this modified version displays a peak in the probability $\Theta_{m}$, which no longer is a monotonically increasing function. The maximum point of \eqref{mTheta} is located at the zero of
\begin{align}
\frac{1}{2}K_{2,2}(\lambda) = K_{2,3}(\lambda),\quad\textnormal{ where }\quad K_{m,n}(\lambda) := \sum_{i=1}^{N}\frac{i^{-m\alpha}}{\left[1+\frac{\lambda}{\lambda_{c}}\frac{i^{-\alpha}}{Z_{\alpha}}\left(\psi_{\alpha}(\lambda)-\phi_{\alpha}(\lambda)\right)\right]^{n}}.
\label{maxmaq}
\end{align}
This new behavior is illustrated in figure \ref{mhmffigure}, and also seen in \cite{KDH23} for a modified version of the heterogneneous mean-field model. 

\section{Simulations}

To compare results from our annealed models with a quenched counterpart (the quenched mean-field model), we conducted computer simulations to solve \eqref{pre_basic_aqmfext} and its modified form \eqref{mme_stationary}, where the interaction matrix $p_{ij}$ was generated by a network with a power-law degree distribution. We first selected a size $N$ and mean degree $c$, then generated a scale-free network with exponent $\gamma$ using the method described in \cite{B20}. Note that the mean degree fixes the number of edges (and hence total degree) in the network, while also determining the minimum degree $m$ (in our case, we chose to express $m$ as a function of $N$, $c$, and $\gamma$). Once the adjacency matrix is constructed, the system of equations \eqref{pre_basic_aqmfext} can be solved to derive the key quantities.

\begin{figure}
\centering
\includegraphics[width=368pt]{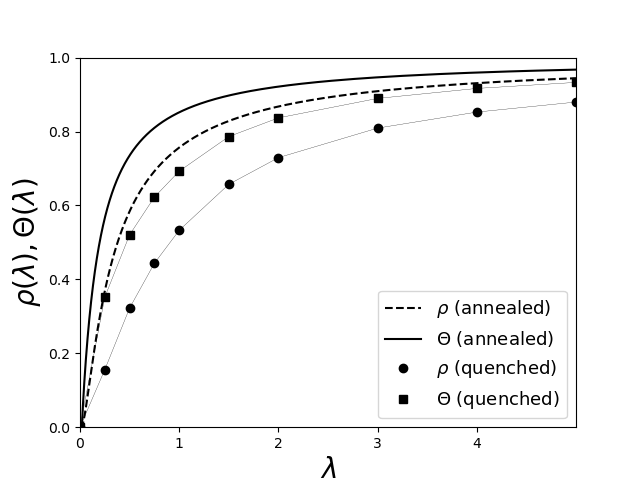}
\caption{
Graph $\rho \times \lambda$ and $\Theta \times \lambda$ for $N=10^{4}$ vertices, with a mean degree of $c=3$, minimum degree $m=2$, and $\gamma = \nicefrac{5}{2}$ ($\alpha = \nicefrac{2}{3}$), using quenched interaction matrices (averaged over $100$ samples). For comparison, the curve associated with the annealed cases is also shown. The error bars are smaller than the marker dimensions. Lines connecting the markers are included solely to aid visualization and do not represent actual data. }
\label{sim_aq}
\end{figure}

We now compare these results with the simulations. It is important to note that we are not examining identical models; unlike in previous sections, where the annealed adjacency matrix was used, the simulations here involve models with quenched interactions. In Figure \ref{sim_aq}, we compare the quenched mean-field model with its annealed version, as discussed in section \ref{subsec:sparse}. We observed a monotonic relationship between prevalence and probability $\Theta$ with respect to the infection rate $\lambda$, and we found that values associated with the annealed case were higher than those for the quenched case. This indicates that annealed interactions propagate more efficiently across the network, whereas quenched interactions create a genuinely sparse interaction network. To illustrate this point, consider the disconnected network from section \ref{subsec:disconnected} for simplicity. If two vertices, $u$ and $v$, are connected, the adjacency matrix entry is $\tilde p_{uv}=1$ in the quenched formulation, whereas in the annealed case, $\tilde p_{uv}<1$ (see \eqref{pij}). Conversely, any two nodes $x$ and $y$ are connected by a nonzero $\tilde{p}_{xy}$ in the annealed case, while this does not hold in the quenched case. Thus, annealed networks are more effective in exploiting the system's connectivity \cite{HM08}.

\begin{figure}
\centering
\includegraphics[width=368pt]{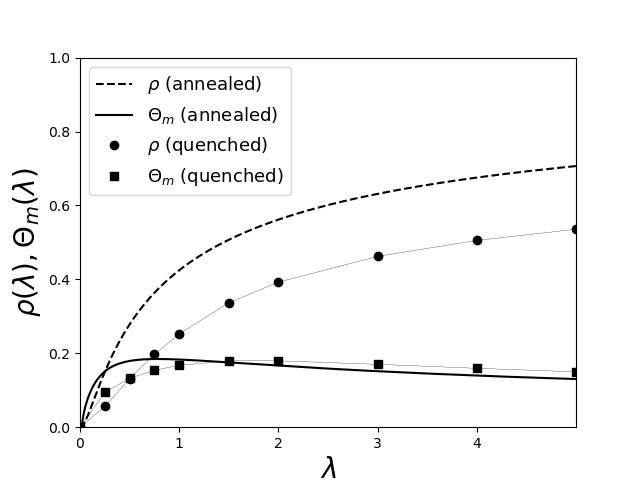}
\caption{
Graph $\rho \times \lambda$ and $\Theta_{m} \times \lambda$ for $N=10^{4}$ vertices, with a mean degree of $c=3$, minimum degree $m=2$, and $\gamma = \nicefrac{5}{2}$ ($\alpha = \nicefrac{2}{3}$), using quenched interaction matrices in the modified version from section \ref{sec:maqmf} (averaged over $100$ samples). For comparison, the curve associated with the annealed cases (see figure \ref{mhmffigure}) is also shown. The error bars are smaller than the marker dimensions. Lines connecting the markers are included solely to aid visualization and do not represent actual data. }
\label{sim_aqm}
\end{figure}

We also found that the modified model analyzed in section \ref{sec:maqmf} exhibits behavior similar to its quenched counterpart, as shown in Figure \ref{sim_aqm}: the prevalence increases monotonically with the infection rate, while the probability $\Theta_{m}$ displays a peak. As with Figure \ref{sim_aq}, prevalence is higher in the annealed case, though no such effect is observed for the probability $\Theta_{m}$ as a consequence of the nonlinear structure introduced in it.

\section{Discussion and Conclusions}

In this work, we introduced an annealed version of the quenched mean-field model. Traditionally, the distinction between quenched and annealed disorder has been crucial in magnetic systems \cite{B59}, where both the states of model constituents (such as spins) and the configurations associated with their interactions play a key role in defining thermodynamic properties. This concept has direct analogies in other fields, including mathematical modeling in epidemiology. In the present work, the states of individuals (susceptible/infected) correspond to the states of particles in magnetic systems, while interactions are represented by an adjacency matrix that characterizes connections between vertices in a graph. As in magnetic systems, the timescale plays a major role in describing the phenomena of interest, highlighting two extreme cases. In the quenched disorder case, the interaction configuration between individuals remains fixed (frozen) during the dynamics. Conversely, in the annealed case, changes in the interaction network between individuals occur much faster than the dynamical processes, allowing us to replace actual connections with an effective mean quantity. Intermediate cases naturally exist, where the evolution of the interaction network occurs on a timescale comparable to the underlying dynamics. This can result from voluntary changes in interactions, environmental shifts, agent mobility, among other factors, and several approaches have been developed to address this kind of problem \cite{CCHHB21, C19, SF18, BC92, dSF15, YTW17, dOSA22}.

In the present work, we introduce the annealed version of a model through its adjacency matrix, ensuring that the underlying network becomes scale-free. Once the structure of contacts between individuals is established, the same approach can be applied. Consequently, this scheme can be extended to other models, such as the susceptible-infected-susceptible (SIR) model \cite{M14, KM27}.

Let us now revisit the objectives and achievements of this work. First, we aim to examine the impact of applying the annealed approximation to an epidemic model originally defined in its quenched form. This approach does not impose a partitioning of nodes into classes where vertices with the same degree are statistically equivalent, as in the HMF approximation. Instead, we obtain the time evolution for each individual node, rather than for classes of nodes with identical degrees. Moreover, this resulting model is also amenable to analytical exploration. We analytically characterize both the critical point and the prevalence near it, demonstrating that both the prevalence and the probability $\Theta$ - the probability of a link encountering an infected node - increase monotonically with the infection rate. This behavior aligns with findings observed in the HMF model \cite{KDH23}.

Secondly, we investigate an epidemic model in which the state of vertices influences individual behavior. Specifically, we focus on scenarios where infected nodes may reduce contact with others due to, for example, voluntary social isolation and/or hospitalization. When we apply a modification inspired by Verhulst’s classical work \cite{V38} to simulate the reduced influence of infected nodes by introducing a mitigation factor to the probability $\Theta$, we observe a qualitative change from the model without this factor. While the infection threshold and the universality class remain the same, the modified probability $\Theta_{m}$ now exhibits a peak and is no longer monotonically increasing. Nevertheless, prevalence continues to increase monotonically with the infection rate. This phenomenon was also observed in \cite{KDH23} within the HMF model defined on a Barabási-Albert network.

We then conducted simulations to compare our results on annealed networks with those on quenched networks. The outcomes show qualitative similarity across all cases: both prevalence and $\Theta$ increase monotonically with the infection rate, and $\Theta_{m}$ exhibits a peak. Moreover, $\Theta_{m}$ obeys \eqref{Thetam<1/4}, as expected. As discussed in the previous section, annealed adjacency matrices can effectively access more nodes. Consequently, the prevalence and probability $\Theta$ - which are linear combinations with positive coefficients of $\rho_{i}$ values ($i\in{1,\ldots,N}$) - are higher for annealed models than for quenched models in our study. However, this observation does not extend to $\Theta_{m}$, which lacks both a linear and increasing relation with the $\rho_{i}$ values.


\section{Acknowledgements}

The authors thank A. S. da Mata for fruitful observations and C. Castellano for helpful comments.


\section{Declarations}

Conflict of interests: the authors declare no competing interests.


\end{document}